# A Cross-disciplinary Framework for the Description of Contextually Mediated Change

Liane Gabora
University of British Columbia

and

Diederik Aerts
Brussels Free University

Address for correspondence regarding the manuscript:

Liane Gabora
Department of Psychology
University of British Columbia
Okanagan Campus, 3333 University Way
Kelowna BC, V1V 1V7, CANADA
liane.gabora[at]ubc.ca
Ph: (250) 807-9849 Fax: (250) 807-8001

**Abstract**
We present a mathematical framework (referred to as Context-driven Actualization of Potential, or CAP) for describing how entities change over time under the influence of a context. The approach facilitates comparison of change of state of entities studied in different disciplines. Processes are seen to differ according to the degree of nondeterminism, and the degree to which they are sensitive to, internalize, and depend upon a particular context. Our analysis suggests that the dynamical evolution of a quantum entity described by the Schrödinger equation is not fundamentally different from change provoked by a measurement often referred to as collapse but a limiting case, with only one way *to* collapse. The biological transition to coded replication is seen as a means of preserving structure in the face of context, and sexual replication as a means of increasing potentiality thus enhancing diversity through interaction with context. The framework sheds light on concepts like selection and fitness, reveals how exceptional Darwinian evolution is as a means of 'change of state', and clarifies in what sense culture (and the creative process underlying it) are Darwinian.

## 1. Introduction

This paper elaborates a general theory of change of state (a nascent, nonmathematical draft of which is presented elsewhere[1]) with the goal of uniting physical, biological, and cultural evolution, not reductively, but through a process that may be referred to as inter-level theorizing (see[2]). Other unifying theories have been put forward, such as that of Treur[3] which takes temporal factorization as its unifying principle. Our scheme focuses on the role of *context*; i.e. on the fact that what entities of different kinds have in common is that they change through a reiterated process of *interaction* with a context. We refer to this fundamental process of change of state under the influence of a context as context-driven actualization of potential, or *CAP*. By "potential" we do not mean determined or preordained; indeed, different forms of evolution differ according to the degree of nondeterminism, as well as degree of contextuality and retention of context-driven change. We conclude with several examples of how the scheme has already born fruit: it suggests a unifying scheme for the two kinds of change in quantum mechanics, [4] illustrates why the earliest forms of life could not have evolved through natural selection,[5] and helps clarify how the concept of evolution applies to culture[6,7] and creative thought[8,9].

## 2. Deterministic and Nondeterministic Evolution under the Influence of a Context

In this section we discuss the kinds of change that must be incoroporated in a general scheme for the description of the evolution of an entity under the influence of a changing or unchanging context. We use the term evolution to mean simply 'change of state'. Thus our neither implies natural selection nor change in the absence of a measurement; the term is thus used as it was prior to both Darwin and Schrödinger.

Since we do not always have perfect knowledge of the state of the entity, the context, and the interaction between them, a general description of an evolutionary process must be able to cope with nondeterminism. Processes differ with respect to the degree of determinism involved in the changes of state that the entity undergoes. Consider an entity—whether it be physical, biological, mental, or some other sort—in a state $p(t_i)$ at an instant of time $t_i$. If it is under the influence of a context $e(t_i)$, and we know with certainty that $p(t_i)$ changes to state $p(t_{i+1})$ at time $t_{i+1}$, we refer to the change of state as *deterministic*. Newtonian physics provides the classic example of deterministic change of state. Knowing the speed and position of a material object, one can predict its speed and position at any time in the future. In many situations, however, an entity in a state $p(t_i)$ at time $t_i$ under the influence of a context $e(t_i)$ may change to any state in the set $\{p_1(t_{i+1}), p_2(t_{i+1}),..., p_n(t_{i+1}),...\}$. When more than one change of state is possible, the process is *nondeterministic*.

Nondeterministic change can be divided into two kinds. In the first, the nondeterminism originates from a lack of knowledge concerning the state of the entity $p(t_i)$ itself. This means that deep down the change is deterministic, but since we lack knowledge about what happens at this deeper level, and since we want to make a model of what we know, the model we make is nondeterministic. This kind of nondeterminism is modeled by a stochastic theory that makes use of a probability structure that satisfies Kolmogorov's axioms.

Another possibility is that the nondeterminism arises through lack of knowledge concerning the context $e(t_i)$, or how that context *interacts* with the entity. Yet another possibility is that the nondeterminism is ontological *i.e*. the universe is at bottom intrinsically nondeterministic. It has been proven that in these cases, the stochastic model to describe this situation necessitates a non-Kolmogorovian probability model. A Kolmogorovian probability model (such as is used in population genetics) cannot be used.[10,11,12,13,14,15] It is only possible to ignore the problem of incomplete knowledge of context if all contexts are equally likely, or if context has a temporary or limited effect. Because the entity has the potential to change to many different states (given the various possible states the context could be in, since we lack precise knowledge of it), we say that it is in a *potentiality state* with respect to context. This is schematically depicted in Figure 1.

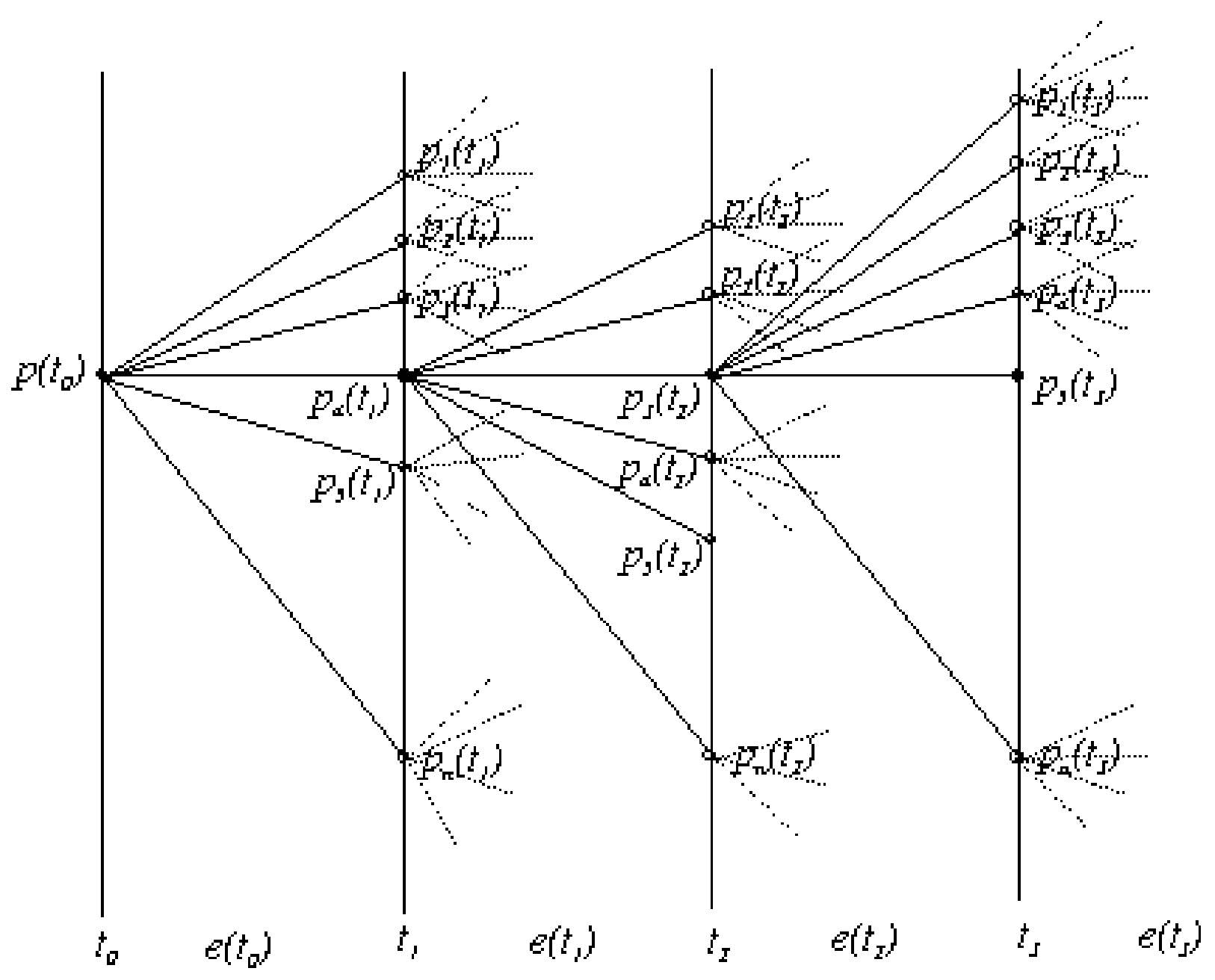

Figure 1. Graphical representation of a general evolution process. Contexts $e(t_0)$, $e(t_1)$, $e(t_2)$ and $e(t_3)$ at times $t_0, t_1, t_2$, and $t_3$, are represented by vertical lines. States of the entity are represented by circles on vertical lines. At time $t_0$ the entity is in state $p(t_0)$. Under the influence of context $e(t_0)$, its state can change to one of the states in the set $\{p_1(t_2), p_2(t_2), p_3(t_2), ..., p_n(t_2), ...\}$. These potential changes are represented by thin lines. Only one change actually takes place, the one represented by a thick line, *i.e.* $p(t_0)$ changes to $p_4(t_1)$. At time $t_1$ the entity in state $p_4(t_1)$ is under the influence of another context $e(t_1)$, and can change to one of $\{p_1(t_2), p_2(t_2), p_3(t_2), ..., p_n(t_2), ...\}$. Again only one change occurs, *i.e.* $p_4(t_1)$ changes to $p_3(t_2)$. The process then starts all over again. Under the influence of a new context $e(t_2)$, the entity can change to one of $\{p_1(t_3), p_2(t_3), p_3(t_3), p_4(t_3), ..., p_n(t_3), ...\}$. Again only one change happens: $p_3(t_2)$ changes to $p_5(t_3)$. The dashed lines from states that have not been actualized at a certain instant indicate that much more potentiality is present at time $t_0$ than explicitly shown. For example, if $p(t_0)$ had changed to $p_2(t_1)$ instead of $p_4(t_2)$, which was possible at time $t_0$, then context $e(t_1)$ would have exerted a different effect on the entity at time $t_1$, such that a new vertical line at time $t_1$ would have to be drawn, representing another pattern of change.

We stress that a potentiality state is not *predetermined*, just waiting for its time to come along, at least not insofar as our models can discern, possibly because we cannot precisely specify the context that will come along and actualize it. Note also that a state is only a potentiality state *in relation to* a certain (incompletely specified) context. It is possible for a state to be a potentiality state with respect to one context, and a deterministic state with respect to another. More precisely, a state that is deterministic with respect to a context can be considered a limit case of a potentiality state, with zero potentiality.

In reality the universe is so complex we can never describe with complete certainty and accuracy the context to which an entity is exposed, and how it interacts with the entity. There is always some possibility of even very unlikely outcomes. However, there are situations in which we can predict the values of relevant variables with sufficient accuracy that we may consider the entity to be in a particular state, and other situations in which there is enough uncertainty to necessitate the concept of potentiality. Thus a formalism for describing the evolution of these entities must take into account the *degree of knowledge* we as observers have about the context.

## 3. SCOP: A Mathematics that Incorporates Nondeterministic Change

We have seen that a description of the evolutionary trajectory of an entity may involve nondeterminism with respect to the state of the entity, the context, or how they interact. An important step toward the development of a general framework for how entities evolve is to find a mathematical structure that can incorporate all these possibilities. There exists an elaborate mathematical framework for describing the change and actualization of potentiality through contextual interaction that was developed for quantum mechanics. However it has several limitations, including the linearity of the Hilbert space, and the fact that one can only describe the extreme case where change of state is *maximally* contextual. Other mathematical theories, such as State COntext Property (SCOP) systems, lift the quantum formalism out of its specific structural limitations, making it possible to describe nondeterministic effects of context in other domains.[16,17,18,19,20,21,22,23,24,25,26,27,28,29] The original motivation for these generalized formalisms was theoretical (as opposed to the need to describe the reality revealed by experiments). In the SCOP formalism, for instance, an entity is described by three sets $\Sigma$, $\mathcal{M}$ and $\mathcal{L}$ and two functions $\mu$ and $\nu$. $\Sigma$ represents that set of states of the entity, $\mathcal{M}$ the set of contexts, and $\mathcal{L}$ the set of properties of the entity.

The function $\mu$ is a probability function that describes how state $p \in \Sigma$ under the influence of context $e \in \mathcal{M}$ changes to state $q \in \Sigma$. Mathematically, this means that $\mu$ is a function defined as follows

$$\begin{aligned} \mu : \Sigma \times \mathcal{M} \times \Sigma &\to [0,1] \\ (q, e, p) &\mapsto \mu(q, e, p) \end{aligned} \qquad (1.1)$$

where $\mu(q, e, p)$ is the probability that the entity in state $p$ changes to state $q$ under the influence of context $e$. Hence $\mu$ describes the structure of the contextual interaction of the entity under study.

### *3.1. Sensitivity to Context*

A parameter that differentiates evolving entities is the degree of *sensitivity to context*, or more precisely, the degree to which a change of state of context evokes a change of state of the entity. Degree of sensitivity to context depends on both the state of the entity and the state of the context. The degree of sensitivity to context is expressed by the probability of collapse. If the probability is close to 1, it means that this context almost with certainty (deterministically) causes the state of the entity to change to the collapsed state. If the probability is close to zero, it means that this context is unlikely to causes the state of the entity to change to the collapsed state. Specifically, if $\mu(q, e, p) = 0$, then $e$ has no influence on the entity in state $p$, and if $\mu(q,e,p)=1$, then $e$ has a strong, deterministic influence on the entity in state $p$. A value of $\mu(q,e,p)$ between 0 and 1, which is the general situation, quantifies the influence of context $e$ on the entity in state $p$.

### *3.2. Degree of Nondeterminism*

Suppose we consider the set

$$\{\mu(q,e,p) \mid e \in M\} \subset [0,1] \tag{1.2}$$

which is a set of points between 0 and 1. Suppose this set equals the singleton $\{1\}$. This would mean that for all context $e \in M$ the entity in state $p$ changed to the entity in state $q$. Hence this would indicate a situation of 'deterministic change independent of context' for the entity, between state $p$ and state $q$. On the contrary, if this set equals the singleton $\{0\}$, this would mean that the entity in state $p$ never changes to state $q$, again independent of context. The general situation of the set being a subset of the interval $[0,1]$ hence models in a detailed way the overal contextual way two states $p$ and $q$ are dynamically connected.

### *3.3. Weights of Properties*

We need a means of expressing that certain properties are more applicable to some entities than others, or more applicable to entities when they are in one state than when they are in another state. The function $\nu$ describes the weight (which is the renormalization of the applicability) of a certain property given a specific state. This means that $\nu$ is a function from the set $\Sigma \times \mathsf{L}$ to the interval $[0,1]$, where $\nu(p,a)$ is the weight of property $a$ for the entity in state $p$. We write

$$\begin{aligned} \nu : \Sigma \times \mathsf{L} &\to [0,1] \\ (p,a) &\mapsto \nu(p,a) \end{aligned} \tag{1.3}$$

The function $\nu$, contrary to $\mu$, describes the internal structure of the entity under investigation. Again we can look to some special situations to clarify how $\nu$ models the internal structure. Suppose that we have $\nu(a,p)=1$. This means that for the entity in state $p$ the property $a$ has weight equal to 1, which means that it is actual with certainty (probability equal to 1). Hence, this represents a situation where the entity, in state $p$, 'has' the property 'in acto'. On the contrary, if $\nu(a,p)=0$, it means that for the entity in state $p$ is not at all actual, but completely potential. A value of $\nu(a,p)$ between 0 and 1 describes in a refined way 'how property $a$ is' when the entity is in state $p$. Hence the set

$$\{\nu(a,p) \mid a \in L\} \subset [0,1] \tag{1.4}$$

gives a good description of the internal structure of the entity, i.e. how properties are depending on the different states the entity can be in.

## 4. Expressing Dynamics: Context-driven Actualization of Potential (CAP)

Context-driven Actualization of Potential, or CAP, is the dynamics of entities modeled by SCOP. This which means that the mathematical model for CAP will be as follows: at moment $t_i$ we have a SCOP

$$(\Sigma, M, L, \mu, \nu)(t_i) = (\Sigma(t_i), M(t_i), L, \mu, \nu) \tag{1.5}$$

where $\Sigma(t_i)$ is the set of states of the entity at time $t_i$ and $M(t_i)$ is the set of relevant contexts at time $t_i$. $L$ is independent of time, since it is the collection of properties of the entity under consideration.

Properties themselves do not change over time, but their status of 'actual' or 'potential' can change with the change of the state of the entity. One should in fact look to it the other way around: a property is an element of the entity idependent of time, or, it is exactly the elements independent of time that give rise to properties. That is also the reason that $L$ and $\nu$ describe the internal context independent structure of the entity under consideration.

In Figure 1 a typical dynamical pattern of CAP is presented. Four consecutive moments of time $t_0, t_1, t_2$ and $t_3$ are considered. It can be seen in the figure how, for example context $e(t_0)$ has an influence on the entity in state $p(t_0)$ which is such that the entity can change into one of the states of the set $\{p_1(t_1), p_2(t_1), p_3(t_1), ..., p_n(t_1)\}$. This is an example of a general nondeterministic type of change. Similar types of changes are represented in the figure, under influence of contexts $e(t_1)$, $e(t_2)$ and $e(t_3)$. What is important to remark is that the actual change taking place is a path through the graph of the figure, but the states not touched by this path remain of influence for the overall pattern of change, since they are potentiality states.

The states $p_0(t_0), p_{i_1}(t_1), p_{i_2}(t_2), ..., p_{i_n}(t_n), ...$ constitute the trajectory of the entity through state space, and describe its evolution in time. Thus, the general evolution process is broadly construed as the incremental change that results from recursive, context-driven actualization of potential, or CAP. A model of an evolutionary process may consist of both *deterministic segments*, where the entity changes state in a way that follows predictably given its previous state and/or the context to which it is exposed, and/or *nondeterministic segments,* where this is not the case. With a generalized quantum formalism such as SCOP it is possible to describe situations with *any* degree of contextuality. In fact, classical and quantum come out as special cases: quantum at one extreme of complete contextuality, and classical at the other extreme, complete lack of contextuality.[30,31,32] This is why it lends itself to the description of context-driven evolution.

#### *4.1. Degree to which Context-driven Change is Retained*

We saw that if $\mu(q, e, p) = 1$, then $e$ has a strong, deterministic influence on the entity in state $p$. However, an entity may be sensitive—readily undergo change of state due to context—but through regulatory mechanisms or self-replication have a tendency to return to a previous state. Thus although it is susceptible to undergoing a change of state from $p$ to $q$, its internal dymaics are such that it goes back to state $p$. A simple example of a situation where this is *not* the case, i.e. context-driven change is retained is a rock breaking in two. An example where it *is* the case, i.e. context-driven change is not retained is the healing of an injury or the birth of offspring with none of the characteristics their parents acquired over their lifetimes. In the case of biological organisms, we are considering two interrelated entities, one embedded in the other: the organism itself, and its lineage.

#### *4.2. Context Independence*

The extent to which a change of context threatens the survival of the entity can be referred to as *context dependence.* The degree to which an entity is able to withstand, not just its *particular* environment, but *any* environment, can be referred to as *context independence*. Sensitivity to and retention of context can lead, in the long run, to *either* context dependence or context independence. This can depend on the variability of the contexts to which an entity is exposed. A static, impoverished environment may provide contexts that foster specializations tailored to *that* particular environment, whereas a dynamic, rich, diverse environment may foster general coping mechanisms. Thus for example, a species that develops an intestine specialized for the absorption of nutrients from a certain plant that is abundant in its environment exhibits context dependence, whereas a species that becomes increasingly more able to consume *any* sort of vegetation exhibits context independence.

Whether an entity exhibits context dependence or independence may simply reflect what one chooses to define as the entity of interest. If an entity splits into multiple 'versions' of itself (as through reproduction) each of which adapts to a different context and thus becomes more context dependent, when all versions are considered different lineages of one *joint* entity, that joint entity is becoming more context-*independent*. Thus for example, while different mammalian species are becoming more context dependent, the kindom as a whole is becoming more context independent.

## 5. Ways the Universe has Found of Actualizing Potential

We now look at how different kinds of evolution fit into the above framework, and how their trajectories differ with respect to the parameters introduced in the previous section. They are all means of actualizing potential that existed due to the state of the entity, the context, and the nature of their interaction, but differ widely with respect to these parameters.

### *5.1. The Evolution of Physical Objects and Particles*

We begin by examining three kinds of change undergone by physical entities. The first is the collapse of quantum particles under the influence of a measurement. The second is the evolution of quantum particles when they are not measured. The third is the change of state of macroscopic physical objects.

*Nondeterministic Collapse of a Quantum Particle*

Let us begin with change of state in the most micro of all domains, quantum mechanics. The central mathematical object in quantum mechanics is a complex Hilbert space $H$, which is a vector space over the field **C** of complex numbers. Vectors $|p(t_i)\rangle$ of this Hilbert space of unit length represent states $p(t_i)$ of a quantum particle. This is expressed in Hilbert space by the bra-ket of the vector with itself being equal to 1, hence

$$\langle p(t_i)|p(t_i)\rangle = 1 \qquad (1.6)$$

A measurement $e(t_i)$ on a quantum particle is described by a self-adjoint operator $E(t_i)$, which is a linear function on the Hilbert space, hence

$$E(t_i): H \rightarrow H \qquad (1.7)$$

such that

$$E(t_i)(a|p_1(t_i)\rangle + b|p_2(t_i)\rangle) = aE(t_i)|p_1(t_i)\rangle + bE(t_i)|p_2(t_i)\rangle \qquad (1.8)$$

A self-adjoint operator always has a set of special states associated with it, the *eigenstates*. A state $p(t_i)$ described in the Hilbert space $H$ by means of the vector $|p(t_i)\rangle$ is an eigenstate of the measurement $e(t_i)$ represented by the self-adjoint operator $E(t_i)$ if and only if we have

$$E(t_i)|p(t_i)\rangle = a|p(t_i)\rangle \qquad (1.9)$$

where a $\in$ **C** is a real number, and $a$ is called the eigenvalue of the measurement $e(t_i)$ the quantum particle being in state $p(t_i)$. Let us denote the set of eigenstates corresponding to the measurement $e(t_i)$ by $\{p_1(t_i), p_2(t_i), ..., p_n(t_i), ...\}$, and the set of their corresponding eigenvalues by $\{a_1, a_2, ..., a_n, ...\}$. Let us mention for mathematical completeness, that only when the self-adjoint operator has a point spectrum do we encounter the above situation in quantum mechanics. Measurements described by operators with a continuous spectrum must be treated in a more sophisticated way. However, this is of no relevance to the points made here.

An eigenstate $p_j(t_i)$ does not change under the influence of the measurement $e(t_i)$ described by the self-adjoint operator $E(t_i)$, and the corresponding eigenvalue $a_j$ is the value obtained deterministically by the measurement. However the set of eigenstates corresponding to a measurement $e(t_i)$ is only a subset of the total set of states of the quantum particle. States that are not eigenstates of the measurement $e(t_i)$ are called superposition states with respect to the measurement $e(t_i)$. Suppose that $q(t_i)$ is such a superposition state of the quantum particle. The vector $|q(t_i)\rangle$ of the Hilbert space representing this superposition state can then always be written as a linear combination, i.e. a superposition, of the vectors $\{|p_1(t_i)\rangle, |p_2(t_i)\rangle, ..., |p_n(t_i)\rangle, ...\}$ representing the eigenstates of the quantum particle. More specifically

$$|q(t_i)\rangle = \sum_j \lambda_j |p_j(t_i)\rangle \tag{1.10}$$

where $\lambda_j \in \mathbf{C}$ are complex numbers such that

$$\sum_j |\lambda_j|^2 = 1 \tag{1.11}$$

The physical meaning of the coefficients $\lambda_j$ is the following: $|\lambda_j|^2$ is the probability that the measurement $e(t_i)$ will yield the value $a_j$ and as a consequence of performing the measurement $e(t_i)$, the superposition state $|q(t_i)\rangle$ will change (collapse) to the eigenstate $|p_j(t_i)\rangle$ with probability equal to $|\lambda_j|^2$. This change of state from a superposition state to an eigenstate is referred to as *collapse*. In general (depending on how many $|\lambda_j|^2$ are non-zero), many eigenstates are possible states to collapse to under the influence of this measurement. In other words, the collapse is non-deterministic. This means that a genuine superposition state is a state of potentiality with respect to the measurement. This suggests that what we refer to as a context is the same thing as what in the standard quantum case is referred to as a measurement.

Thus a quantum entity exists in general in a superposition state, and a measurement causes it to collapse nondeterministically to an eigenstate of that measurement. The specifics of the measurement constitute the context that elicit one of the states that were previously potential. Its evolution cannot be examined without performing measurements—that is, introducing contexts—but the contexts unavoidably affect its evolution. The evolution of a quantum particle is an extreme case of nondeterministic change, as well as of context sensitivity, because its state at any point in time reflects the context to which it is exposed.

*Evolution of Quantum Particles*

The other mode of change in standard quantum mechanics is the dynamical change of state when no measurement is executed, referred to as 'evolution'. This evolution is described by the Schrödinger equation, and it is considered a fundamentally different kind of change from 'collapse' under the influence of a measurement. The Schrödinger evolution is described by the Hamiltonian $H$ which is the self-adjoint operator corresponding to the measurement $h(t_i)$ of the energy of the quantum particle, and the Schrödinger equation

$$|p(t+t_i)\rangle = e^{\frac{-iHt}{\hbar}} |p(t_i)\rangle \tag{1.12}$$

where $|p(t+t_i)\rangle$ is the vector representing the state $p(t+t_i)$ of the quantum particle at time $t+t_i$ and $|p(t_i)\rangle$ is the vector representing the state $p(t_i)$ of the quantum particle at time $t_i$, while $e^{\frac{-iHt}{\hbar}}$ is the unitary operator describing the time translation of state $p(t_i)$ to state $p(t+t_i)$, and $\hbar$ is Planck's constant. It is possible to interpret the change of the state of a quantum entity as described by the Schrödinger equation as an effect of context, namely a context that is the rest of the universe. All the fields and matter present in the rest of the universe contribute to the effect. The change is deterministic. Specifically, if the quantum entity at a certain time $t_i$ is in state $p(t_i)$, and the only change that takes place is this dynamical change governed by the Schrödinger equation, then state $p(t+t_i)$ at any time $t+t_i$ later than $t_i$ is determined.

For historical reasons, physicists think of a measurement not as a context, but as a process that gives rise to outcomes that are read off a measurement apparatus. In this scheme of thought, the simplest measurements are assumed to be those with two possible outcomes. A measurement with one outcome is rightly not thought of as a measurement, because if the same outcome always occurs, nothing has been compared and/or measured. However, when measurements are construed as contexts, we see that the measurement with two possible outcomes is not the simplest change possible. It is the deterministic evolution process—which can be conceived as a measurement with one outcome—that is the simplest kind of change. This means that in quantum mechanics the effect of context on change is as follows:

- When the context is the rest of the universe, its influence on the state of a quantum entity is deterministic, as described by the Schrödinger equation.
- When the context is a measurement, its influence on a genuine *superposition state* is *nondeterministic,* described as a process of collapse.
- When the context is a measurement, its influence on an *eigenstate* is *deterministic*, the eigenstate is not changed by the measurement.

Thus, under the CAP framework, the two basic processes of change in quantum mechanics are united; what has been referred to as dynamical evolution is not fundamentally different from collapse. They are both processes of actualization of potentiality under the influence of a context. In dynamical evolution there is only one possible outcome, thus it is deterministic. In collapse, until the state of the entity becomes an eigenstate, there is more than one possible outcome, thus it is nondeterministic.

It was mentioned that the standard quantum formalism contains some fundamental structural restrictions. For example, the state space, i.e. the complex Hilbert space, is a linear space. It was also mentioned that more general axiomatic approaches to quantum mechanica have been developed where these structural restrictions are not present. More specifically, the state space in such a quantum axiomatic approach does not need to be a linear space. The SCOP approach which is the underlying mathematical framework for the 'context driven actualization of potential' change envisaged here, is such an axiomatic quantum 'and' classical approach, i.e. both standard quantum mechanics and classical mechanics are special cases of this SCOP formalism.[33,34,35,36,37,38,39,40,41] There is an even more specific structural restriction of standard quantum mechanics, namely the fact that the Schrödinger equation is a linear equation, and can only represent changes of states described by a unitary transformation. For this reason, even staying within the standard quantum mechanical Hilbert space formalism for the state space, non-linear and hence non-unitary evolution more general than the Schrödinger one have been considered and studied on many occasions.[42,43,44,45] All these variations on standard quantum mechanics are special cases of the SCOP formalism, and hence CAP incorporates the types of changes modeled by them.

*Evolution of Classical Physical Entities*
Classical physical entities are the paradigmatic example of lack of sensitivity to and internalization of context, and deterministic change of state. However, theorists are continually expanding their models to include more of the context surrounding an entity in order to better predict its behavior, which suggests that things are not so tidy in the world of classical physical objects as Newtonian physics suggests. Macro-level physical entities may exhibit structure similar to the entanglement of quantum mechanics.[46] This is the case when change of state of the entity cannot be predicted due to lack of knowledge of how it interacts with its context.[47,48]

### *5.2. Biological Evolution*
Some theorists seeking to develop an integrative framework for the physical and life sciences (e.g.[49]) focus on tools or phenomena that are applicable to or appear in both such as power laws and adaptive landscapes. The real challenge, however, is to develop an adaptive framework that encompasses what is unique about biological life, what makes some matter alive, and what give rise to lineages that undergo adaptive modification. This is closely tied with the origins of structure with the capacity to self-replicate, and the refinement of this capacity through the emergence of the genetic code.[50,51] Thus, this section is divided into four parts. The first concerns autocatalytic sets of polymers, which replicated themselves (sloppily) without genetic information, and are therefore are widely believed to be the first forms that can be considered 'alive'. The second concerns organisms after genetically coded replication was established but prior to sexual reproduction. The third concerns sexually reproducing organisms. The fourth part is a general comment on the structure of biochemical change.

*The Earliest Life Forms*
Early life forms were more sensitive to context and prone to internalize context than present-day life because their replication took place not according to instructions (such as a genetic code), but through happenstance interactions. In Kauffman's model of the origin of life,[52] polymers catalyze reactions that generate other polymers, increasing their joint complexity, until together as a whole they form something that can more or less replicate itself.[53] The set is *autocatalytically closed* because although no polymer catalyzes its own replication, each catalyzes the replication of another member of the set. So long as each polymer is getting duplicated somewhere in the set, eventually multiple copies of all polymers exist. Thus it can self-replicate, and continue to do so indefinitely, or at least until it changes so drastically that its self-replicating structure breaks down. (Notice that 'death' of such life forms is not a particularly noticeable event; the only difference between a dead organism and an alive one is that the alive one continues to spawn new replicants.) Replication is far from perfect, so 'offspring' are unlikely to be identical to their 'parent'. Different chance encounters of polymers, or differences in their relative concentrations, or the appearance of new polymers, could all result in different polymers catalyzing a given reaction, which in turn altered the set of reactions to be catalyzed. Context was readily internalized by incorporating elements of the environment, thus there was plenty of room for heritable variation.

Recent work has been shown that the dynamical structure of biochemical reactions is quantum-like, above and beyond their microscopic (and obvious) quantum structure.[54,55,56] This stems from the fact that not only in the micro-domain where standard quantum structures are known to exist, but also in other domains where change-of-state depends on how an entity interacts with its context, the resulting probabilities can be non-Kolmogorovian, and the appropriate formalisms for describing them are either the quantum formalisms or mathematical generalizations of them. Kolmogorovian probability models consider only actualized entities, and their actualized interactions, and assumes that all evolutionary change is steered by these actualized entities, and their actualized interactions.Within CAP, potential states of entities, and potential interactions between them, can be described.

*Genetic Code Impedes Retention of Context in Lineage*
A significant transition in the history of life was the transition from uncoded, self-organized replication to replication as per instructions given by a genetic code. This has been beautifully described and modelled by Vetsigian, Goldenfeld, and Woese[57] They refer to the transition from sloppy self-replication of autocatalytic form to precise replication using a genetic code as the Darwinian transition, since it is at this point that traits acquired at the level of *individuals* were no longer inherited and natural selection, a *population-level* mechanism of change, becomes applicable. We saw that prior to coded replication, a change to one polymer would still be present in offspring after budding occurred, and this could cause other changes that have a significant effect on the lineage further downstream. There was nothing to prohibit inheritance of acquired characteristics. But with the advent of explicit self-assembly instructions, acquired characteristics were no longer passed on to the next generation. The reason for this stems from the fact that, as first noted by von Neumann[58], they are self-replicating automata, meaning that they contain a coded instruction set that is used in two distinct ways: (1) as a self-assembly code that is *interpreted* during *development*, and (2) as self-instructions that are *passively copied* during *reproduction*. It is this separation of how the code is used to generate offspring, and how the code is used during development, that is at the foundation of their lack of inheritance of acquired traits.

Since acquired characteristics were no longer being passed on to the next generation, the process became more constrained, robust, and shielded from external influence. (Thus for example, if one cuts off the tail of a mouse, its offspring will have tails of a normal length.) A context-driven change of state of an organism only affects its lineage if it impacts the generation and survival of progeny (such as by affecting the capacity to attract mates, or engage in parental care). Clearly, the transition from uncoded to coded replication, while ensuring fidelity of replication, decreased long-term sensitivity to and internalization of context, and thus capacity for context independence. Since one generation was almost certainly identical to the next, the evolution became more deterministic. As a result, in comparison with entities of other sorts, biological entities are resistant to internalization and retention of context-driven change. Though the term 'adaptation' is most closely associated with biology, biological form is resistant to adaptation. This explains why it has been possible to develop a theory of biological evolution that all but ignores the problem of incomplete knowledge of context. As we saw earlier, it is possible to ignore this problem if all contexts are equally likely, or if context has a limited effect on heritability. In biology, since acquired traits are not heritable, the only contextual interactions that exert much of an effect are those that affect the generation of offspring. Thus it is because classical stochastic models work fine when lack of knowledge concerns the state of the entity and not the context that natural selection has for so long been viewed as adequate for the description of biological evolution. In Aerts, Czachor, and D'Hooghe[59] and Aerts et al.[60], Darwinian evolution is analyzed with respect to potentiality using a specific example, and various possible (and speculative) consequences of this difference are put forward.

*Sexual Reproduction*
With the advent of sexual reproduction, the contextuality of biological evolution increased. Consider an organism that is heterozygous for trait X with two alleles *A* and *a*. The potential of this *Aa* organism gets actualized differently depending on the context provided by the genotype of the organism's mate. In the context of an *AA* mate, the *Aa* organism's potential is constrained to include only *AA* or *Aa* offspring. In the context of an *aa* mate, it has the potential for *Aa* or *aa* offspring, and once again some of this potential might get actualized. And so forth. But while the mate *constrains* the organism's potential, the mate is necessary to *actualize* some of this potential in the form of offspring. In other words, the genome of the mate simultaneously makes *some* aspects of the *Aa* organism's potentiality possible, and *others* impossible. An organism exists in a state of potentiality with respect to the different offspring (variants of itself) it could produce with a particular mate. In other words, a mate constitutes a context for which an organism is a state of potentiality. One can get away with ignoring this to the extent that one can assume mating is random. Note that since a species is delineated according to the capacity of individuals to mate with one another, speciation can be viewed as the situation wherein one variant no longer has the potential to create a context for the other for which its state is a potentiality state with respect to offspring. A species can be said to be adapted to the extent that its previous states *could have* collapsed to different outcomes in different contexts, and thus to the extent its form reflects the particular contexts to which it *was* exposed. Note also that although over time species becomes increasingly context dependent, collectively they are becoming more context *in*dependent. (For virtually any ecological niche there exists *some* branch of life that can cope with it.)

Some (*e.g.*[61]) argue for expansion of the concept of selection to other hierarchical levels, *e.g*. group selection. We agree with Kitcher[62] that 'despite the vast amount of ink lavished upon the idea of "higher-order" processes', once we have the causal story, it's a matter of convention whether we say that selection is operating at the level of the species, the organism, the genotype, or the gene. It is not the concept of selection that needs expansion, but the embedding of selection in a framework for how change can occur. The actual is but the realized fragment of the potential, and selection works *only* on this fragment, what is already actual. We can now return to our question about what natural selection has to say about the fitness of the offspring you might have with one mate as opposed to another. The answer is of course, nothing, but why? Because the situation involves actualization of potential and nondeterminism with respect to context, and as we have seen, a nonclassical formalism is necessary to describe the change of state involved. The CAP perspective also clarifies why fitness has been so hard to nail down. We agree with Krimbas[63] that fitness is a property of neither organism nor environment, but emerges at the interface between them, and changes from case to case.

### *5.3. Change of State in Cognitive Processes*

Campbell[64,65,66,67] argues that a stream of creative thought is a Darwinian process. The basic idea is that we generate new ideas through variation and selection: 'mutate' the current thought in a multitude of different ways, select the variant that looks best, mutate *it* in various ways and select the best, and so forth, until a satisfactory idea results. Thus a stream of thought is treated as a series of tiny selections. This view has been extended, most notably by Simonton.[68,69,70,71] The problems with this thesis are outlined in detail elsewhere[72],[73], but one that is evident following our preceding discussion is that thoughts simply are not von Neumann self-replicating automata. Another is that selection theory requires multiple, distinct, simultaneously-actualized states. But in cognition, each thought or cognitive state changes the 'selection pressure' against which the next is evaluated; they are not simultaneously selected amongst. The mind can exist in a state of potentiality, and change through interaction with the context to a state that is genuinely new, not just an element of a pre-existing set of states. Creative thought is a matter of honing in on an idea by redescribing successive iterations of it from different real or imagined perspectives[74]; *i.e.* actualizing potential through exposure to different contexts. Once again, the description of contextual change of state introduces a non-Kolmogorovian probability distribution, and a classical formalism such as selection theory cannot be used. Thus an idea certainly changes as it gets mulled over in a stream of thought, and indeed it appears to evolve, but the process is not Darwinian. Campbell's error is to treat a set of *potential*, contextually elicited states of *one* entity as if they were *actual* states of a *collection* of entities, or possible states with no effect of context, even though the mathematical structure of the two situations is completely different. In a stream of thought, neither are all contexts equally likely, nor does context have a limited effect on future iterations. So the assumptions that make classical stochastic models useful approximations do not hold.

### *5.4. Cultural Evolution*

Culture, even more often than creative thought, is interpretted in evolutionary terms. Some scientists view culture merely as a contributing to the biological evolution of our species. Increasingly, however, culture is viewd as an evolutionary process in and of its own (though one that still is intertwined with, and influences, biological evolution). In this section we look at how cultural evolution (as a second evolutionary process) fits into the CAP framework.

*Culture Evolves Without a Self-Assembly Code*

The basic unit of culture has been assumed to be the behavior or artifact, or the mental representations or ideas that give rise to concrete cultural forms. The meme notion further implies that these cultural units are replicators,[75] misleadingly, because it does not consist of self-assembly instructions.[76] (An idea may *retain* structure as it passes from one individual to another, but does not *replicate* it.) Looking at cultural evolution from the CAP framework we ask: what is *really* changing through cultural processes? Because of the distributed nature of human memory, it is never is just one discrete 'meme'affected by a cultural experience; it is ones view of how the world hangs together, ones' model of reality, or *worldview*. A worldview is not merely a collection of discrete ideas or memes (nor do ideas or memes form an interlocking set like puzzle pieces) because each context impacts it differently; concepts and ideas are always colored by the situation in which they are evoked.[77,78,79] Indeed it has been argued that a worldview is a replicator.[80] We saw that living organisms prior to the genetic code—a pre-RNA set of autocatalytic polymers—were *primitive replicators* because they generate self-similar structure, but in a self-organized, emergent, piecemeal manner, eventually, for each polymer, there existed another that catalyzed its formation. Since there was no self-assembly instructions to copy from, there was no explicit copying going on. The presence of a given catalytic polymer, say X, simply speeded up the rate at which certain reactions took place, while another polymer, say Y, influenced the reaction that generated X. Just as polymers catalyze reactions that generate other polymers, retrieval of an item from memory can trigger another, which triggers yet another and so forth, thereby cross-linking memories, ideas, and so forth into a conceptual web. Like the autocatalytic sets of poymers considered earlier,[81] the result can be described as a connected closure structure.[82,83] Elements of a worldview are regenerated through social learning. Since as with Kauffman's origin of life scenario the process occurrs in a self-organized, piecemeal autocatalytic manner, through bottom-up interactions rather than a top-down code, worldviews like the earliest life forms replicate with low fidelity, and their evolution is highly nondeterministic.

*Inheritance of Acquired Traits in Culture*

As with the earliest pre-DNA forms of life, characteristics of a worldview acquired over a lifetime are heritable. We hear a joke and, in sharing it with others, give it our own slant. We create a disco version of Beethoven's Fifth Symphony and a rap version of that. The evolutionary trajectory of a worldview makes itself known indirectly via the behavior and artifacts it manifests under the influence of the contexts it encounters. (For example, when you explain to a child how to brush ones' teeth, certain facets of your worldview are revealed, while your writing of a poem reveals other facets.)

Because acquired traits are heritable in culture, the probability of splitting into multiple variants is high. These variants can range from virtually identical to virtually impossible to trace back to the same 'parent' idea. They affect, and are affected by, the minds that encounter them. For example, books can affect all the individuals who read them, and these individuals subsequently provide new contexts for the possible further evolution of the ideas they described and stories they told.

## 6. Conclusions

This paper introduced a general framework for characterizing how entities evolve through context-driven actualization of potential (CAP). By this we mean an entity has the *potential* to change in different ways under different contexts. Some aspects of this potentiality are actualized when the entity undergoes a change of state through *interaction with the particular context* it encounters. The interaction between entity and context may also change the context, and the constraints and affordances it offers the entity. Thus the entity undergoes another change of state, and so forth, recursively.

When evolution is construed as the incremental change that results from recursive, context-driven actualization of potential, the domains through which we have carved up reality can be united under one umbrella. Quantum, classical, biological, cognitive, and cultural evolution appear as different ways in which potential that is present due to the state of an entity, its context, and the nature of their interaction. They differ according to the degree of sensitivity to context, internalization of context, dependence upon a particular context. nondeterminism due to lack of knowledge concerning the state of the entity, and nondeterminism due to lack of knowledge concerning the state of the context.

The reason potentiality and contextuality are so important stems from the fact that we inevitably have incomplete knowledge of the universe in which an entity is operating. When the state of the entity of interest and/or context are in constant flux, or undergoing change at a resolution below that which we can detect but nevertheless affect what emerges at the entity-context interface, this gives rise in a natural way to nondeterministic change. Nondeterminism that arises through lack of knowledge concerning the state of the entity can be described by classical stochastic models (Markov processes) because the probability structure is Kolmogorovian. However, nondeterminism that arises through lack of knowledge concerning the interaction between entity and context introduces a non-Kolmogorovian probability model[84] on the state space, necessitating a nonclassical formalism. Historically, the first nonclassical formalism was the quantum formalism. This formalism has since been generalized to describe situations involving nonlinearity, and varying degrees of contextuality.

Let us sum up a few of the more interesting results to come out of this framework. It has been thought that the two modes of change in quantum mechanics—dynamical evolution of the quantum entity as per the Schrödinger equation, and the collapse that takes place when the quantum entity is measured—were fundamentally different. However, when the measurement is seen to be a context, we notice that it is always a context that could actualize the potential of the entity in different ways. Indeed, if one knows the outcome with certainty one does not perform a measurement; it is only when there is more than one possible value that a measurement is performed. Thus the two modes of change in quantum mechanics are united; the dynamical evolution of a quantum entity as per the Schrödinger equation reduces to a collapse for which there was only one way *to* collapse (i.e. only one possible outcome), hence deterministic collapse. This also holds for the deterministic evolution of classical entities. This constitutes a true paradigm shift, for evolution and collapse have been thought to be two fundamentally different processes.

Looking at biological evolution from the CAP perspective, self-replication appears as a means of testing the integrity of an entity—or rather different versions of an entity—against different contexts. While individuals and even species become increasingly context-dependent, the joint entity of living organisms becomes increasingly context-*in*dependent.The genetic code afforded primitive life protection against contextually-induced disintegration of self-replication capacity, at the cost of decreased diversity. The onset of sexual reproduction increased potentiality, and thus possible trajectories for biological form. The CAP framework supports the notion that fitness is a property of neither organism nor environment, but emerges at the interface between them. The concept of potential fitness includes all possible evolutionary trajectories under all possible contexts. Since it involves nondeterminism with respect to context, unless context has a limited effect or all possible contexts are equally likely, a nonclassical formalism is necessary to describe the novel form that results when an organism interacts with its environment in a way that makes some of its potential became actual (where actual fitness refers only to the *realized* segment of its potentiality). It now becomes clear why natural selection has been able to tell us much about changes in frequencies of existing forms, but little about how new forms emerge in the first place! The CAP framework provides a perspective from which we can see why the neo-Darwinian view of evolution has been satisfactory for so long, and it wasn't until after other processes become prominently viewed in evolutionary terms that the time was ripe for potentiality and contextuality to be taken seriously. We also see how unique the genetic code is, and the consequent lack of retention of context-driven change. The effects of contextual interaction in biology are in the long-run largely invisible; context affects lineages only by influencing the number and nature of offspring. Natural selection is such an exceptional means of change, it is no wonder it does not transfer readily to other domains. Note that it is often said that because acquired traits are inherited in culture, culture should not be viewed in evolutionary terms. It is ironic that this critique also applies to the earliest stage of biological evolution itself. What was true of early life is also true of the replication of worldviews: acquired characteristics can be inherited. Modern life is unique in this sense.

The same argument holds for what happens in a stream of creative thought. The mathematical formulation of the theory of natural selection requires that in any given iteration there be multiple distinct, actualized states. In cognition however, each successive mental state changes the context in which the next is evaluated; they are not simultaneously selected amongst. Creative thought is a matter of honing in on an idea by redescribing successive iterations of it from different real or imagined perspectives; actualizing potential through exposure to different contexts. Thus selection theory is not applicable to the formal description of a stream of thought, and to the extent that creative thought powers cultural change, it is of limited applicability there as well. Once again, a nonclassical formalism is necessary.

The notion of culture as a Darwinian process probably derives from the fact that the means through which a creative mind manifests itself in the world—language, art, and so forth—exist as discrete entities such as stories and paintings. This can lead to the assumption that discrete creative artifacts in the world spring forth from corresponding discrete, pre-formed entities in the brain. This in turn leads to the assumption that novelty gets generated through that most celebrated of all change-generating mechanisms, Darwinian selection, and that ideas and artifacts must therefore be replicators. However, an idea or artifact is not a replicator because it does not consist of coded self-assembly instructions, and thus does not make copies of itself. Moreover, ideas and artifacts do not arise out of separate, distinct compartments in the brain, but emerge from a dynamically and contextually modifiable, web-like memory structure, a melting pot in which different components continually merge and blend, get experienced in new ways as they arise in new contexts and combinations. The CAP perspective suggests instead that the basic unit and the replicator of culture is an integrated network of knowledge, attitudes, ideas, and so forth; that is, an internal model of the world, or worldview, and that ideas and artifacts are how a worldview reveals itself under a particular context.

**Acknowledgments**

We would like to thank Ping Ao for comments on the manuscript. This research was supported by Grant G.0339.02 of the Flemish Fund for Scientific Research and a research grant from Foundation for the Future.

[25] Mackey G. *Math. Found. Quant. Mech.* (John Benjamin, Reading, MA, 1963).
[26] Piron C. *Found. Quant. Phys.* (W.A. Benjamin, Reading, MA, 1976).
[27] Piron C. *Helvetica Physica Acta* 62(82), 439-68 (1989).
[28] Piron C. *Mécanique Quantique: Bases et Applications* (Press Polytechnique de Lausanne, Lausanne, Switzerland, 1990)
[29] Randall C. & Foulis D. 'The Operational Approach to Quantum Mechanics', in C. Hooker (ed.), *Physical Theories as Logico-Operational Structures* 167 (Reidel, Dordrecht, The Netherlands, 1978).
[30] Aerts D. 'Being and change foundations of a realistic operational formalism'. In D. Aerts, M. Czachor & T. Durt (eds.), *Probing the Structure of Quantum Mechanics: Nonlinearity, Nonlocality, Probability and Axiomatics* 71-110. (World Scientific, Singapore, 2002).
[31] Piron C. *Found. Quant. Phys.* (W.A. Benjamin, Reading, MA, 1976).
[32] Aerts D. *J. Math. Phys.* 24, 2441-2454 (1983).
[33] Aerts D. 'Being and change foundations of a realistic operational formalism'. In D. Aerts, M. Czachor & T. Durt (eds.), *Probing the Structure of Quantum Mechanics: Nonlinearity, Nonlocality, Probability and Axiomatics* 71-110 (World Scientific, Singapore, 2002).
[34] Randall C. & Foulis D. 'A Mathematical Setting for Inductive Reasoning', in C. Hooker (ed.), *Foundations of Probability Theory, Statistical Inference, and Statistical Theories of Science III*, 169 (Reidel, Dordrecht, The Netherlands, 1976)
[35] Aerts D. & Durt T. *Found. Phys.* 24, 1353-1368 (1994).
[36] Foulis D. & Randall C. 'What are Quantum Logics and what Ought they to Be?', in E. Beltrametti & B. van Fraassen (eds.), *Current Issues in Quantum Logic* 35 (Plenum, New York, 1981).
[37] Piron C. *Found. Quant. Phys.* (W.A. Benjamin, Reading, MA, 1976).
[38] Piron C. *Helvetica Physica Acta* 62(82), 439-68 (1989).
[39] Piron C. *Mécanique Quantique: Bases et Applications* (Press Polytechnique de Lausanne, Lausanne, Switzerland, 1990).
[40] Randall C. & Foulis D. 'The Operational Approach to Quantum Mechanics', in C. Hooker (ed.), *Physical Theories as Logico-Operational Structures* 167 (Reidel, Dordrecht, The Netherlands, 1978).
[41] Aerts D. *J. Math. Phys.* 24, 2441-2454 (1983).
[42] Gisin N. *Phys. Lett. A* 143, 1 (1990).
[43] Mielnik B. *Phys. Lett. A* 289, 1-8 (2001).
[44] Polchinski J. *Phys. Rev. Lett.* 66, 379-400 (1991).
[45] Weinberg S. *Ann. Phys.* 194, 336-386 (1989).
[46] Aerts D., Aerts S., Broekaert J. & Gabora L. *Found. Phys.* 30, 1387–1414 (2000).
[47] Aerts D. 'Being and change foundations of a realistic operational formalism'. In D. Aerts, M. Czachor & T. Durt (eds.), *Probing the Structure of Quantum Mechanics: Nonlinearity, Nonlocality, Probability and Axiomatics* 71-110 (World Scientific, Singapore, 2002).
[48] Aerts D. *J. Math. Phys.* 24, 2441-2454 (1983).
[49] Ao P. Laws of Darwinian evolutionary theory. *Phys. Life Rev.* 2, 117-156 (2005).
[50] Langton C. G. In C. Langton, C. Taylor, J. D. Farmer & S. Rasmussen (eds.) *Artificial Life II* (Addison-Wesley, Redwood City, CA, 1992).
[51] Schrödinger E. *What Is Life? The Physical Aspect of the Living Cell* (Cambridge University Press, Cambridge, 1944).
[52] Kauffman S. *Origins of Order* (Oxford University Press, New York, 1993).
[53] The reason this works is because when polymers interact, the number of different polymers increases exponentially, but the number of reactions by which they can interconvert increases faster than their total number. Thus, as their diversity increases, so does the probability that some subset of the total reaches a critical point where there is a catalytic pathway to every member.
[54] Aerts D., Czachor M., Gabora L. & Polk P. *AIP Conference Proceedings* 810, 19-33. (2006).
[55] Aerts D., Czachor M., Gabora L., Kuna M., Posiewnik A., Pykacz J. & Syty M. *Phys. Rev. E* 67, 051926 (2003).
[56] Aerts D. & Czachor M. *Nonlinearity* 19, 575-589 (2006).
[57] Vetsigian, K., Woese, C., and Goldenfeld, N. *Proceedings of the National Academy Science USA* 103: 10696-10701 (2006).
[58] Von Neumann J. *The Theory of Self-Reproducing Automata* (University of Illinois Press, Chicago, 1966).
[59] Aerts D., Czachor M. & D'Hooghe B. 'Towards a quantum evolutionary scheme: violating Bell's inequalities in language'. In N. Gontier, J. P. Van Bendegem & D. Aerts (eds.), *Evolutionary Epistemology, Language and Culture - A non adaptationist systems theoretical approach* [Theory and Decision Library Series A: Philosophy and Methodology of

the Social Sciences. Series editor: Julian Nida-Ruemelin] (Dordrecht, Springer, 2006).

[60] Aerts D., Bundervoet S., Czachor M., D'Hooghe B., Gabora L. & Polk P. 'On the foundations of the theory of evolution'. In M. Locker (ed.), *Systems Theory in Philosophy and Religion. Vols. I&II* (IIAS, Windsor, Ont, 2007).

[61] Gould S. J. *Bully for Brontosaurus: Reflections in Natural History* 63-66 (W.W. Norton & Company, New York, 1991).

[62] Kitcher P. *Bio. & Phil.* 19(1), 1-15 (2004).

[63] Krimbas C.B. *Bio. & Phil.* 19(2), 185-204 (2004).

[64] Campbell D.T. *Psych. Rev.* 67, 380-400 (1960).

[65] Campbell D.T. 'Variation and selective retention in socio-cultural evolution'. In H.R. Barringer, G I. Blanksten & R.W. Mack (eds), *Social change in developing areas: A reinterpretation of evolutionary theory* (Cambridge, Schenkman, 1965).

[66] Campbell D.T. 'Evolutionary epistemology'. In P.A. Schilp (ed.), *The Philosophy of Karl Popper* (Open Court, LaSalle, IL, 1974).

[67] Campbell D.T. 'Evolutionary epistemology'. In G. Radnitzky & W.W. Bartley III (eds.) *Evolutionary epistomology: Rationality, and the sociology of knowledge* (Open Court, LaSalle, IL, 1987).

[68] Simonton D.K. *J. Creat. Behav.* 32(3), 153-58 (1998).

[69] Simonton D.K. *Origins of Genius: Darwinian Perspectives on Creativity* (Oxford, New York, 1999).

[70] Simonton D.K. 'Creativity as Blind Variation and Selective Retention: Is the Creative Process Darwinian?', *Psychological Inquiry* 10, 309-28. 1999b,

[71] Simonton D.K. *Creat. Res. J.* in press (2007).

[72] Gabora L. *J. Creat. Behav*. 39(4), 65-87 (2005).

[73] Gabora, L. *Creativity Research J.* (2007a).

[74] Gabora, L. *Creativity Research J.* (2007b).

[75] Dawkins R. *The Selfish Gene* (Oxford University Press, Oxford, 1976).

[76] Gabora L. *Bio. & Phil.* 19(1), 127-143 (2004).

[77] Barsalou L. *Mem. & Cog.* 10, 82-93 (1982).

[78] Gabora L. & Aerts D. 'Contextualizing concepts'. In *Proceedings of the 15th International FLAIRS Conference, Pensacola Florida, May 14-17, 2002, AAAI* (2002).

[79] Gabora L. & Aerts D. *J. Experi. & Theor. AI* 14(4), 327-358 (2002).

[80] Gabora L. *Bio. & Phil.* 19(1), 127-143 (2004).

[81] Kauffman S. *Origins of order* (Oxford University Press, New York, 1993).

[82] Gabora L. Psycoloquy, 9(67) (1998).

[83] Gabora L. 'Conceptual closure: Weaving memories into an interconnected worldview'. In G. Van de Vijver & J. Chandler (eds.), *Closure: Emergent Organizations and their Dynamics. Ann. NY Acad. Sci.* 901, 42-53 (2000).

[84] For example Bayes' formula for conditional probability is not satisfied.